\documentclass[a4paper]{article}

\usepackage[hyperfootnotes=false]{hyperref}
\hypersetup{nolinks=true}
\usepackage{INTERSPEECH_v2}
\usepackage{float}
\usepackage{float}
\usepackage{amsmath}
\usepackage{graphicx}
\usepackage{algorithm} 
\usepackage{algorithmic}
\usepackage[justification=centering]{caption}
\usepackage[colorinlistoftodos]{todonotes}

\usepackage{pbox}
\usepackage{multirow}
\usepackage{makecell}

\usepackage{setspace}

\title{Efficient minimum word error rate training of RNN-Transducer for end-to-end speech recognition}

\name{Jinxi~Guo, Gautam~Tiwari, Jasha~Droppo, Maarten~Van~Segbroeck, \\ Che-Wei~Huang, Andreas~Stolcke, Roland~Maas}
\address{Amazon.com, USA}
\email{\{jinxiguo, tgautam, drojasha, segbrm, cheweh, stolcke, rmaas\}@amazon.com}

\begin{document}

\maketitle
\begin{abstract}
In this work, we propose a novel and efficient minimum word error rate (MWER) training method for RNN-Transducer (RNN-T). Unlike previous work on this topic, which performs on-the-fly limited-size beam-search decoding and generates alignment scores for expected edit-distance computation, in our proposed method, we re-calculate and sum scores of all the possible alignments for each hypothesis in N-best lists. The hypothesis probability scores and back-propagated gradients are calculated efficiently using the forward-backward algorithm. Moreover, the proposed method allows us to decouple the decoding and training processes, and thus we can perform offline parallel-decoding and MWER training for each subset iteratively. Experimental results show that this proposed semi-on-the-fly method can speed up the on-the-fly method by 6 times and result in a similar WER improvement (3.6\%) over a baseline RNN-T model. The proposed MWER training can also effectively reduce high-deletion errors (9.2\% WER-reduction) introduced by RNN-T models when EOS is added for endpointer. Further improvement can be achieved if we use a proposed RNN-T rescoring method to re-rank hypotheses and use external RNN-LM to perform additional rescoring. The best system achieves a 5\% relative improvement on an English test-set of real far-field recordings and a 11.6\% WER reduction on music-domain utterances.

\end{abstract}

\section{Introduction}

End-to-end models for automatic speech recognition (ASR) have gained popularity in recent years as a way to fold separate components of a conventional ASR system (i.e., acoustic, pronunciation and language models) into a single neural network. Examples of such models include connectionist temporal classification (CTC) based models \cite{graves2006connectionist}, recurrent neural network transducer (RNN-T) \cite{graves2012sequence}, and attention-based seq2seq models \cite{chan2016listen}. Among these models, RNN-T is the most suitable streaming end-to-end recognizer, which has shown competitive performance compared to conventional systems \cite{rao2017exploring,he2019streaming}. 

RNN-T models are typically trained with RNN-T loss, which aims to improve the log-likelihood of training data. However, few research work has investigated sequential discriminative training criteria for RNN-T models.  For traditional hybrid system, a state-level minimum Bayes risk (sMBR) training criteria has been successfully applied \cite{vesely2013sequence}. In terms of word-level MBR training, sampling-based approaches were proposed for CTC \cite{graves2014towards,Shannon17} and recurrent neural aligner (RNA) \cite{Sak17} to minimize expected WER. Recently, minimum WER (MWER) training \cite{Rohit18} has been proposed to train attention-based seq2seq models, and shown significant improvements. In the context of RNN-T models, a recent work in \cite{Weng19rnnt} proposed to perform MBR training by making use of the decoded alignments of N-best hypotheses.

In this work, a novel and efficient MWER training method is proposed for RNN-T. Comparing to the existing method in \cite{Weng19rnnt}, which uses relative small beam size to perform on-the-fly decoding to generate alignments scores and N-best list,  we re-calculate the scores of all the possible alignments for each hypothesis in a given N-best list and use the combined scores for MWER training. The hypothesis probability scores and back-propagation gradients are calculated using the forward-backward algorithm similar to RNN-T training, which is very efficient in both speed and memory usage. Since on-the-fly N-best generation is expensive, and given the fact that the N-best lists don't change much within a short training period, we perform offline decoding and MWER training for each subset iteratively, which allows us to speed up both decoding and training process separately. We apply the proposed techniques on large scale far-field English data sets, and we show that the proposed semi-on-the-fly decoding and training method can speed up the MWER training process by 6 times without affecting WER improvement (3.6\%). Our proposed MWER training method can also address the issue of high deletion errors introduced by the RNN-T model when end-of-sentence (EOS) token is used for endpointer, and show a 9.2\% relative improvement. We achieve further improvements by proposing an RNN-T rescoring approach to re-rank the hypotheses and use RNN-LM to rescore the updated N-best list. Our best system achieves 5\% and 11.6\% relative improvements on general and music domain datasets respectively.

\section{Baseline RNN-T model}

\begin{figure}[t]
\centering
\includegraphics[width=0.77\linewidth]{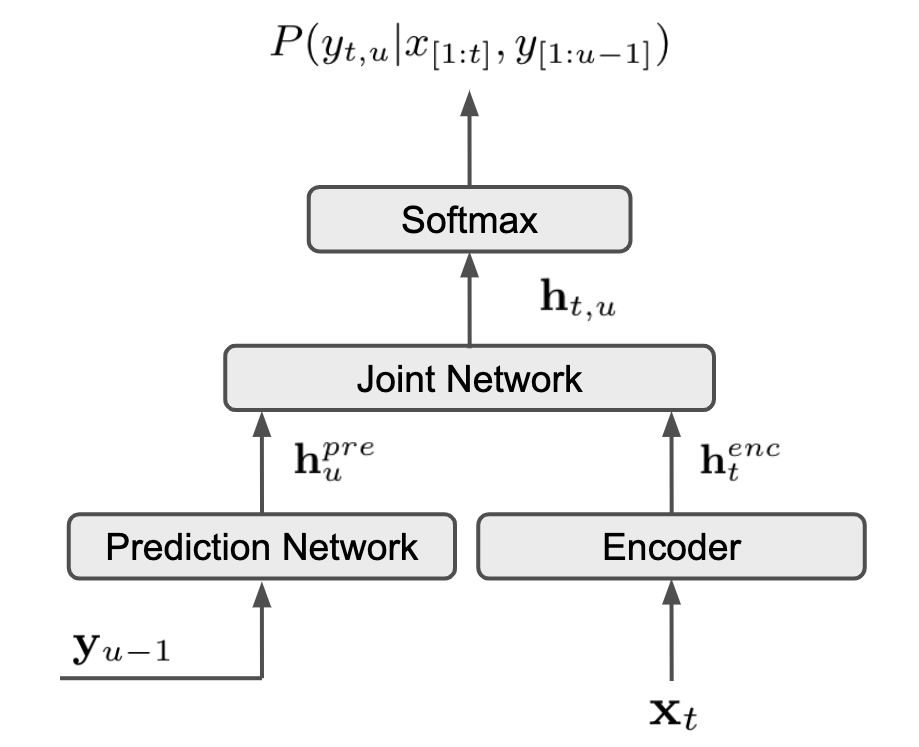}
\caption{RNN-Transducer architecture.}
\label{fig:RNN-T}
\vspace{-3.5mm}
\end{figure}

Figure \ref{fig:RNN-T} shows the architecture for an RNN-T model, which consists of an encoder, a prediction network and a joint network. The encoder is analogous to an acoustic model that receives acoustic features vectors $\mathbf{x}_t$ and converts them into a sequence of hidden states $\mathbf{h}_{t}^{enc}$, where $t$ is the time index. The prediction network acts as a language model that takes previous sub-word label prediction $\mathbf{y}_{u-1}$ as input, and produces hidden representation $\mathbf{h}_{u}^{pre}$, where u is label index. The joint network is usually a feed-forward network that takes each combination of encoder output $\mathbf{h}_{t}^{enc}$ and prediction network output $\mathbf{h}_{u}^{pre}$, and computes output logits $\mathbf{h}_{t,u}$. The final posterior for each output token $P(k|t,u)$ is obtained after applying the softmax operation.

The loss function of RNN-T is the negative log posterior of output label sequence $\mathbf{y}$ given input acoustic feature $\mathbf{x}$: 
\begin{equation}
L_\text{RNN-T} = -\log{P(\mathbf{y}|\mathbf{x})},
\label{eq:1}
\end{equation}
where $P(\mathbf{y}|\mathbf{x}) = \sum_{\hat{\mathbf{y}}}P(\hat{\mathbf{y}}|\mathbf{x})$, $\hat{\mathbf{y}} \in A$. $A$ is the set of all the possible alignments (with blank labels) between input $x$ and output $y$. 

To minimize $L$, an efficient forward-backward algorithm was proposed in \cite{graves2012sequence}. The key part of the algorithm is to calculate the derivatives of loss $L$ with respect to $P(k|t,u)$, as shown below:
%

\begin{equation}
\frac{\partial L_\text{RNN-T}}{\partial P(k|t,u)}= - \frac{\alpha(t,u)}{P(y|x)}
\begin{cases}
\beta(t, u+1)& \text{if}\:k=y_{u+1}\\
\beta(t+1,u)& \text{if}\:k=\varnothing \\
0 & \text{otherwise}
\end{cases}
\label{eq:2}
\end{equation}
where the forward variable $\alpha(t,u)$ is the probability of outputting $y_{[1:u]}$ during $x_{[1:t]}$, and the backward variable $\beta(t,u)$ is the probability of outputting $y_{[u+1:U]}$ during $x_{[t:T]}$. $\varnothing$ represents the blank label.

During inference, the N-best list of the utterance is generated using beam search decoding algorithm as described in \cite{graves2012sequence}, with a minor change in our implementation: during the search process, we skip the summation over prefixes and only sum the probabilities of the beam candidates which correspond to the same hypothesis sequence (contains only non-blank labels).

\section{Efficient minimum word error rate training of RNN-T}

In the following subsections, we will first give a detailed introduction of minimum word error rate training (MWER) for end-to-end models and previous related works. Then we will introduce our proposed novel and efficient MWER training method of RNN-T.

\subsection{Minimum word error rate training}

Minimum word error rate training \cite{Shannon17,Rohit18} is also referred to word-level minimum Bayes risk (MBR) training. Mathematically, given a hypothesis $y$ and the ground-truth sequence $y^r$, we denote the number of word errors in $y$ by $R(y,y^{r})$. In order to directly minimize the word error rate and perform discriminative training between hypotheses, the expected number of word errors over hypotheses was proposed as loss function:

\begin{equation}
L_\text{MWER} = E[R(y,y^{r})] = \sum_{y}P(y|x)R(y,y^{r}).
\label{eq:3}
\end{equation}

Computing all the possible hypotheses is intractable, and therefore N-best lists are commonly used to approximate the expectation in Eq.~\ref{eq:3}, since the probability mass is concentrated on the top-N hypotheses $y_i\in \text{nbest(x)}$ for end-to-end models. Thus Eq.~\ref{eq:3} can be approximated as below:

\begin{equation}
L_\text{MWER} = \sum_{y_{i} \in \text{nbest}(x)}\hat{P}(y_{i}|x)R(y_{i},y^{r}),
\label{eq:4}
\end{equation}
where $\hat{P}(y_{i}|x) = \frac{P(y_{i}|x)}{\sum_{y_{i} \in \text{nbest}(x)} P(y_{i}|x)}$, represents the normalized distribution over N-best hypotheses. We will show that this normalization is critical to perform discriminative training between hypotheses, and reduce the variance of the gradients.

We now introduce our derivation to perform MWER training for end-to-end models (e.g. RNN-T). First of all, we need to derive the gradients of $L_\text{MWER}$ w.r.t the outputs of the model after decoding, i.e. $\log{P(y_{i}|x)}$. Let's reformulate Eq.~\ref{eq:4} using $\log{P(y_{i}|x)}$ as:
\begin{equation}
L_\text{MWER} = \sum_{y_{i} \in \text{nbest}(x)}\text{softmax}(\log{P(y_{i}|x)})R(y_{i},y^{r}).
\label{eq:5}
\end{equation}
Then the derivative can be calculated as,
\begin{equation}
\frac{\partial L_\text{MWER}}{\partial \log{P(y_{i}|x)}} = \sum_{y_{j} \in \text{nbest}(x)}\frac{\partial \text{softmax}(\log{P(y_{j}|x)})}{\partial \log{P(y_{i}|x)}}R(y_{j},y^{r}).
\label{eq:6}
\end{equation}
According to the derivative equation of softmax function and $\hat{P}(y_{i}|x)=\text{softmax}(\log{P(y_{i}|x)})$, we can get:
\begin{equation}
\frac{\partial \text{softmax}(\log{P(y_{j}|x)})}{\partial \log{P(y_{i}|x)}}=
\begin{cases}
\hat{P}(y_{j}|x)(1-\hat{P}(y_{i}|x))  & {j = i}\\
-\hat{P}(y_{j}|x)\hat{P}(y_{i}|x)   & {j \neq i}
\end{cases}
\label{eq:7}
\end{equation}
Therefore, the derivative of $L_\text{MWER}$ can be finally written as:
\begin{eqnarray}
\begin{split}
\frac{\partial L_{\text{MWER}}}{\partial \log{P(y_{i}|x)}} &\!=\! \hat{P}(y_{i}|x)R(y_{i},y^{r})- \\
& \hat{P}(y_{i}|x)\!\sum_{y_{j}\in\text{nbest}(x)}\!\hat{P}(y_{j}|x)R(y_{j},y^{r}) \\
&\!=\! \hat{P}(y_{i}|x)(R(y_{i},y^{r}) - \hat{R}),
\end{split}
\label{eq:8}
\end{eqnarray}
where $\hat{R} = \sum_{y_{j}\in \text{nbest}(x)} \hat{P}(y_{j}|x)R(y_{j},y^{r})$, is the expected number of word errors within the N-best list. Subtracting $\hat{R}$ from $R$ also helps to reduce the variance of gradients. From Eq.~\ref{eq:8}, we can observe that given $\sum_{i}\hat{P}(y_{i}|x)=1$, for hypotheses with fewer errors than $\hat{R}$, $R(y_{i},y^{r}) - \hat{R} < 0$ and $\log{P(y_{i}|x)}$ will be boosted; while the probability of hypotheses with more errors than $\hat{R}$ will be decreased. Compared with RNN-T loss which only aims to increase the probability of the ground-truth transcription, the MWER loss, instead, performs discriminative training between hypotheses.

\subsection{Related research}
In \cite{Rohit18}, the authors first proposed to use MWER loss to train LAS models, which provides significant improvements. However, in terms of RNN-T model, few research studies have been reported. One of the key issues for RNN-T is that each hypothesis in the N-best represents many different alignments generated during beam search decoding. The choice of beam size and decoding strategy determines the number of alignments used to calculate the probability of a hypothesis, while in LAS models, the probability of a unique hypothesis is independent of the chosen beam size.

In \cite{Weng19rnnt}, the authors proposed to use the decoded alignment sequences (with blanks) from the RNN-T model to train a word-level MBR loss. However, due the speed and memory issues introduced by the approach, a beam size of 2 was adopted, which indicated that only a relatively limited number of alignments were used for their MBR training. Similarly, a very recent work \cite{li2020towards} also uses a similar approach.

\subsection{Proposed MWER loss for RNN-T}
We are motivated by the fact that RNN-T loss uses all the possible alignments of a given label sequences to calculate their probabilities, and the forward-backward algorithm can make the training very efficient in both speed and memory usage. Therefore a novel MWER training algorithm is proposed to capture all the alignments for each hypotheses in the N-best list. 

More specifically, instead of calculating $\log{P(y_{i}|x)}$ of each hypothesis during beam-search decoding, we feed the hypothesis $y_i$ back into the prediction network, in order to generate all the possible alignments of $y_i$ given $x$. Therefore, we can directly calculate the derivative of $L_{\text{MWER}}$ w.r.t $P_{y_i}(k|t,u)$ following the chain rule as:

\begin{equation}
\frac{\partial L_{\text{MWER}}}{\partial P_{y_i}(k|t,u)}
\!=\!\frac{\partial L_{\text{MWER}}}{\partial \log{P(y_{i}|x)}} 
\frac{\partial \log{P(y_{i}|x)}}{\partial P_{y_i}(k|t,u)},
\label{eq:9}
\end{equation}
where
\begin{equation}
\frac{\partial L_{\text{MWER}}}{\partial \log{P(y_{i}|x)}}
\!=\! \hat{P}(y_{i}|x)(R(y_{i},y^{r}) - \hat{R}),
\label{eq:10}
\end{equation}

\begin{equation}
\frac{\partial \log{P(y_{i}|x)}}{\partial P_{y_i}(k|t,u)} \!=\! \frac{\alpha_{y_{i}}(t,u)}{P(y_{i}|x)}
\begin{cases}
\beta_{y_{i}}(t, u+1)& \text{if}\:k=y_{i(u+1)}\\
\beta_{y_{i}}(t+1,u)& \text{if}\:k=\varnothing \\
0 & \text{otherwise}
\end{cases}
\label{eq:11}
\end{equation}

where $\alpha$ and $\beta$ follow the same definitions as in Eq.~\ref{eq:2}.

There are several advantages of this proposed approach. First of all, the limitation of the beam size when performing beam search decoding could lead to bias towards top hypotheses and thus miss alignments of other hypotheses. Recalculating all the possible alignments of each hypothesis accurately generates scores and back-propagation paths for gradient calculation. Secondly, when performing the proposed MWER training, it smartly groups alignments of the same hypothesis together and uses the forward-backward algorithm to efficiently calculate gradients.

\subsection{Semi-on-the-fly N-best generation}
On-the-fly N-best generation is expensive, since decoding is performed for each batch. This makes MWER training to be very time consuming. However, we have observed that the N-best list and oracle WER for a given utterance don't change much within a short training period, while the alignments and probability scores are changing after each training step (one batch). Since our proposed MWER training strategy recalculates all the possible alignments and probabilities of hypotheses within each batch anyway, there may be no need to on-the-fly generate and update N-best lists. Therefore, in this paper, we propose a semi-on-the-fly N-best generation approach as shown in Algorithm \ref{alg:1}. 

Within each subset $S_i$ (defined in Algorithm \ref{alg:1}), we perform off-line decoding and N-best list generation. Off-line decoding makes it feasible to perform parallel computing across utterances using large amount of CPU threads, which reduces the decoding time significantly. During the MWER training step, the training speed is as fast as the regular RNN-T model. Therefore, with a sufficient number of CPUs, the overall training speed is in the same order of RNN-T model. 

\begin{algorithm}[h]
\caption{Procedure for MWER training with semi-on-the-fly N-best generation}
\begin{algorithmic}[1]
\STATE Split the training data $S$ into K splits: $S_i$ ($i \in [1,K]$);
\STATE Denote a well-trained RNN-T model as $M_{1,0}$;
\FOR{$i=1$ to N \text{(Num of Epochs)}}
\FOR{$j=1$ to K \text{(Num of Splits)}}
\STATE 1) use current model $M_{i,j-1}$ to decode subset $S_{j}$ with multiple CPU threads; save the decoded N-best lists.
\STATE 2) use subset $S_{j}$ together with saved N-best lists to train the RNN-T model with the proposed MWER loss; after training finished, we get the updated model $M_{i,j}$.
\ENDFOR
\STATE $M_{i+1,0}$ = $M_{i,K}$
\ENDFOR
\end{algorithmic}
\label{alg:1}
\end{algorithm}
\vspace{-3.5mm}

\subsection{RNN-T rescoring}

During MWER training, we feed hypotheses back into the RNN-T prediction network to generate scores; while in decoding, we use the beam-search algorithm to generate scores. Therefore, motivated by the fact that there exists mismatch between training and decoding, we propose a novel 2nd-pass rescoring method: after 1st-pass decoding, we rescore the hypotheses by feeding them into the RNN-T model in order to generate more accurate scores, and re-rank the hypotheses.

\subsection{Language model rescoring}
To incorporate text-only data into end-to-end ASR models, external LMs are commonly used. In this paper, we train an external RNN-LM and use it to rescore the N-best hypotheses generated by the RNN-T models. Specifically, during inference our objective is to find the most likely sub-word sequences given the score from RNN-T model $P(y|x)$ and the LM $P_{LM}(y)$:
\begin{equation}
y^* = \arg\max_{y}(\log P(y|x)+\lambda \log P_{LM}(y)/|y|).
\label{eq:12}
\end{equation}
For $P(y|x)$, we will use the probability scores generated from the baseline RNN-T model, MWER trained RNN-T model and MWER trained RNN-T model after RNN-T rescoring. For LM scores, we use length-normalized probabilities.

\section{Experimental setup}
\subsection{Data sets}
Our RNN-T models are trained on a $\sim$23,000 hour training set consisting of 25 million far-field English utterances for voice control. We also augment the acoustic training data with the SpecAug \cite{park2019specaugment} algorithm, in order to improve the robustness. Our model is evaluated on a test set which contains $\sim$16K far-field English utterances. We will also show results on test sets of specific domains. The text corpus used to train the external LM contains $\sim$55 million English sentences.  

\subsection{Configurations of RNN-T model and RNN-LM}
All experiments use 64-dimensional log-Mel features, computed with 25ms window and 10ms shift. Each feature vector is stacked with 2 frames to the left and downsampled to a 30ms frame rate. For RNN-T, the encoder consists of 5 LSTM layers, where each layer has 1024 hidden units. The prediction network has 2 LSTM layers of 1024 units and an embedding layer of 512 units. For the choice of joint network, we compared a feed-forward network with a simple addition operation as in \cite{graves2012sequence}, which adds the outputs generated from encoder and prediction network directly. Given the GPU memory limitation (16G Nvidia V100), a simple addition operation allows us to train a model with a larger batch size, which provides similar performance and much faster training speed. Therefore, an addition operation is adopted for the experiments in this paper. The softmax layer consists of 10k output units, and is trained to predict wordpiece tokens, which are generated using the byte pair encoding algorithm \cite{schuster2012japanese}. A stacked RNN with two unidirectional LSTM layers is used as an external language model. The embedding dimension is also set to 512. The same 10K wordpiece token set is used as RNN-T. 

All models are trained using the Adam optimizer \cite{KingmaBa15}, with a learning rate schedule including an initial linear warm-up phase, a constant phase, and an exponential decay phase following \cite{Mia18}. For RNN-T training, the learning rates for the constant phase and end of exponential decay phase are 5e-4 and 1e-5, respectively.

\subsection{MWER training set-ups}
We use a converged RNN-T model as the seed model for MWER training. To generate N-best lists, a beam size of 4 is used for all the MWER training experiments in this paper. We will compare the on-the-fly N-best generation method with the proposed semi-on-the-fly method introduced in Section 3.4. The learning rate schedule for MWER training also includes three phases same as RNN-T training. The learning rates for the constant phase and end of exponential decay phase are 1e-5 and 1e-6, respectively. We use a dev set to monitor the MWER loss during training.

\section{Results}
\subsection{Baseline RNN-T model}
For the baseline RNN-T model, we show the decoding results using different beam widths. Previous research \cite{Weng19rnnt} shows that applying temperature into softmax is beneficial during decoding, therefore we also investigate the choice of temperature in Table 1. We have observed that a softmax temperature of 1.2 gives the biggest improvement. Note that,  all WERs reported in this paper are normalized WER numbers, and the number is calculated by $\text{WER}/first(\text{WER})$, where $first(\text{WER})$ is the first WER number reported in this paper (row 2, column 2 in Table 1).

\begin{table} [t!]
\begin{center}
    \begin{tabular}{ | l || c | c | c  | c  |p{5cm}|}
    \hline
    Beam size & 3 & 4 & 8 & 16  \\ \hline
    RNN-T (temp 1.0)  & 1.000 & 0.991 & 0.978 & 0.979   \\ \hline
    RNN-T (temp 1.2) & 0.997 & 0.986 & 0.962 & 0.968   \\ \hline
    \end{tabular}
\end{center}
 \caption{\label{tab:table2} Normalized WERs (calculated by $\text{WER}/first(\text{WER})$, where $first(\text{WER})$ is the first WER number reported in this paper) of baseline RNN-T models when decoding with different temperatures.}
\vspace{-6mm}
\end{table}

\subsection{Improvements from MWER training}
We first present the results of the proposed MWER training method applied on the baseline RNN-T model using on-the-fly N-best generation method in Table 2. Table 2 indicates that a relative 3.6\% improvement can be achieved compared with the baseline model. For the proposed semi-on-the-fly N-best generation approach, we experimented with K = 10 and 20, where each split contains around 2.5M and 1.25M utterances, respectively. When updating N-best lists for every 2.5M utterances, the model converged to a local minimum point where the loss didn't decrease anymore; while, when we reduced the number utterances to 1.25M, the loss converged to a similar point as the on-the-fly model. Table 2 compares the decoding results of these two methods, and the WERs are very similar. In terms of training time, semi-on-the-fly approach has similar training speed as RNN-T model, which is 6 times faster compared with the on-the-fly approach. 


\begin{table} [t]
\begin{center}
    \begin{tabular}{| l || c | c | c  | c  |p{5cm}|}
    \hline
    Beam size & 3 & 4 & 8 & 16  \\ \hline
    RNN-T baseline  & 0.997 & 0.986 & 0.962 & 0.968   \\ \hline
    + MWER (On-the-fly) & 0.965 & 0.956 & 0.939 & 0.935   \\ \hline
    + MWER (Semi-on-the-fly) & 0.965 & 0.954 & 0.939 & 0.933  \\ \hline
    \end{tabular}
\end{center}
\caption{\label{tab:table3} Normalized WERs of MWER-trained models using on-the-fly and semi-on-the-fly strategies. }
\vspace{-8mm}
\end{table}

\subsection{Effects of MWER training on RNN-T with EOS}
The endpointer (EP) model, which decides if a speaker has stop speaking, is important for voice control based ASR model. Besides a commonly used independent EP model, \cite{chang2019joint} proposed to add an end-of-sentence (EOS) token to train RNN-T, in order to use it signaling the end of speech. However, introducing EOS token always results in high deletion errors for RNN-T model due to the early trigger issue as shown in Table 3. MWER training, which makes use of the relative error distances between hypotheses, can effectively penalize early trigger of EOS. In Table 3, we can see a significant improvement (9.2\%) when applying our proposed MWER training method on the RNN-T model with EOS added. This paper only reports WER numbers with the standard decoding set up, and as shown in \cite{chang2019joint,li2020towards}, penalizing EOS during decoding can further help to find a trade off point between WERs and EP latency, which will not be discussed here.


\begin{table} [t]
\begin{center}
    \begin{tabular}{| l || c |p{5cm}|}
    \hline
    Beam size & 16  \\ \hline
    RNN-T baseline  & 0.968  \\ \hline
    RNN-T with EOS & 1.083  \\ \hline
    RNN-T with EOS + MWER & 0.983 \\ \hline
    \end{tabular}
\end{center}
\caption{\label{tab:table3} Normalized WERs when applying MWER training on RNN-T model with EOS. }
\vspace{-4.5mm}
\end{table}

\subsection{2nd-pass rescoring}
Two 2nd-pass rescoring approaches are presented in Table 4, i.e. RNN-T rescoring (proposed in Section 3.5) and RNN-LM rescoring. Experiments E1 and E2 in Table 4 shows that MWER training clearly outperforms LM rescoring when they are applied on the baseline RNN-T model. Comparing E4 and E2, we can observe that RNN-T rescoring provides additional improvement, due to the fact that it matches the decoding with MWER training and generates more accurate scores. The effect of RNN-T rescoring is more prominent for small beam size condition. Based on experiments E3, E4 and  E5, we can conclude that both LM and RNN-T rescoring help to improve the performance of the MWER-trained RNN-T model, and they are complementary with each other. The discriminatively trained RNN-T model can help to select the correct hypothesis when applying LM rescoring on N-best lists. The overall improvement for E5 over the baseline model E0 is around 5\%.
We furthermore analyzed the performance across domains and present results on music-related intents. In Table 4 (last column), we notice that MWER training and rescoring improve the baseline model significantly (11.6\%). This may be due to the fact that MWER training helps to correctly recognize the rare words, like song names.

\begin{table} [t!]
\begin{center}
    \begin{tabular}{ | l || c | c | c   |p{5cm}|}
    \hline
    Domain  & \multicolumn{2}{c|}{General} & Music \\ \hline
    Beam size  & 4 & 16 & 16 \\ \hline
    RNN-T baseline (E0)   & 0.986  & 0.968 & 0.997   \\ \hline
    E0 + LM rescoring (E1) & 0.970  & 0.956 & ---  \\ \hline
    E0 + MWER (E2)& 0.954  & 0.933 & 0.893 \\ \hline
    E2 + LM rescoring (E3)  & 0.944 & 0.925 & ---  \\ \hline
    E2 + RNN-T rescoring (E4) & 0.943  & 0.929 & 0.887   \\ \hline
    E4 + LM rescoring (E5)   & 0.936 & 0.920 & 0.881 \\ \hline
    Rel Imp (E5 vs. E0) & 5 \% & 5\% & 11.6\% \\ \hline
    \end{tabular}
\end{center}
 \caption{\label{tab:table2} Normalized WERs of different set-ups on general domain and music domain test sets. ``Rel Imp" refers to ``Relative Improvement".}
\vspace{-7mm}
\end{table}

\section{Conclusion}
In this paper, we propose a novel and efficient MWER training algorithm of RNN-T based ASR models. To perform MWER training, we sum the scores of all the possible alignments for each hypothesis in the N-best list, and use them to calculate the expected edit distance between reference and hypotheses. The forward-backward algorithm is used to efficiently calculate scores and gradients. Moreover, a semi-on-the-fly decoding and training algorithm is proposed, which can speed up the whole training process by 6 times without degradation. The proposed MWER training can also significantly reduce the high deletion errors when EOS is added for EP. We furthermore introduce a RNN-T rescoring method to re-rank the hypotheses and use an external RNN-LM to perform additional rescoring. Overall, we find that our best system shows 5\% relative improvement on a general domain test set and 11.6\% relative improvement on music domain data.

\section{Acknowledgement}
The authors would like to thank Leif Raedel for help on the RNN-LM training, and thank Simon Wiesler, Jahn Heymann, Ilya Sklyar, Andreas Schwarz, Valentin Mendelev, Egor Lakomkin, Harish Arsikere, Dylan Mckinney and Anirudh Raju for the support on the RNN-T baseline system.

\newpage
\bibliographystyle{IEEEtran}

\bibliography{mybib}


\end{document}